\documentclass[twocolumn,aps,prl,reprint,groupedaddress,showkeys]{revtex4}
\usepackage{graphicx}
\usepackage{color}
\usepackage{amsmath}

\newcommand{\be}{\begin{equation}}
\newcommand{\ee}{\end{equation}}
\newcommand{\bea}{\begin{eqnarray}}
\newcommand{\eea}{\end{eqnarray}}

\newcommand{\la}{\langle}
\newcommand{\ra}{\rangle}

\newcommand{\lp}{\left(}
\newcommand{\rp}{\right)}

\newcommand{\bra}[1]{\la #1|}
\newcommand{\ket}[1]{| #1\ra}
\newcommand{\sgn}{{\rm sgn}\,}

\renewcommand{\vec}[1]{{\bf #1}}
\renewcommand{\hat}[1]{{\widehat #1}}

\begin{document}

\title{Topological Bloch Bands in Graphene Superlattices}
\author{Justin C. W. Song$^{1,2,3}$}
\email{jcwsong@caltech.edu}
\author{Polnop Samutpraphoot$^3$}
\author{Leonid S. Levitov$^3$}
\email{levitov@mit.edu}
 \affiliation{$^1$ Walter Burke Institute of Theoretical Physics, California Institute of Technology, California 91125, USA}
 \affiliation{$^2$ Department of Physics, California Institute of Technology, California 91125, USA}
  \affiliation{$^3$ Department of Physics, Massachusetts Institute of Technology, Cambridge, Massachusetts 02139, USA}
 
\begin{abstract}
We outline an approach to endow a plain vanilla material with topological properties by creating topological bands in stacks of manifestly nontopological atomically thin materials. The approach is illustrated with a model system comprised of graphene stacked atop hexagonal-boron-nitride. In this case, the Berry curvature of the electron Bloch bands is  highly sensitive to the stacking configuration. As a result, electron topology can be controlled by crystal axes alignment, granting a practical route to designer topological materials.
Berry curvature manifests itself in transport via the valley Hall effect and long-range chargeless valley currents.  The non-local electrical response mediated by such currents  provides diagnostics for band topology.
\end{abstract} 
\maketitle

%{\bf Significance Statement} - {\it 
%A family of designer topological materials is introduced, comprised of stacks of two-dimensional materials which by themselves are not topological, such as graphene. Previously, topological bands in graphene were presumed either impossible or impractical. 
%The designer approach turns graphene into 
%a robust platform with which a host of topological behavior can be realized and explored.}

\section{Introduction}

Electronic states in topological materials possess unique properties
including a Hall effect without an applied magnetic field\cite{hasan,dixiao,nagaosa} and topologically protected edge states.\cite{TKNN,halperin} 
Accessing non-trivial electron topology depends on identifying materials in which symmetry and interactions produce topological Bloch bands. Such bands can only arise when multiple requirements, such as a multi-band structure with a Berry phase and suitable symmetry, are fulfilled. As a result, topological bands are found in only a handful of exotic materials in which good transport properties are often lacking. Formulating practical methods for transforming widely available materials with a reasonably high carrier mobility (such as Silicon, or Graphene) into a topological phase remains a grand challenge. 

\begin{figure}[t!]
\includegraphics[width=0.5\textwidth]{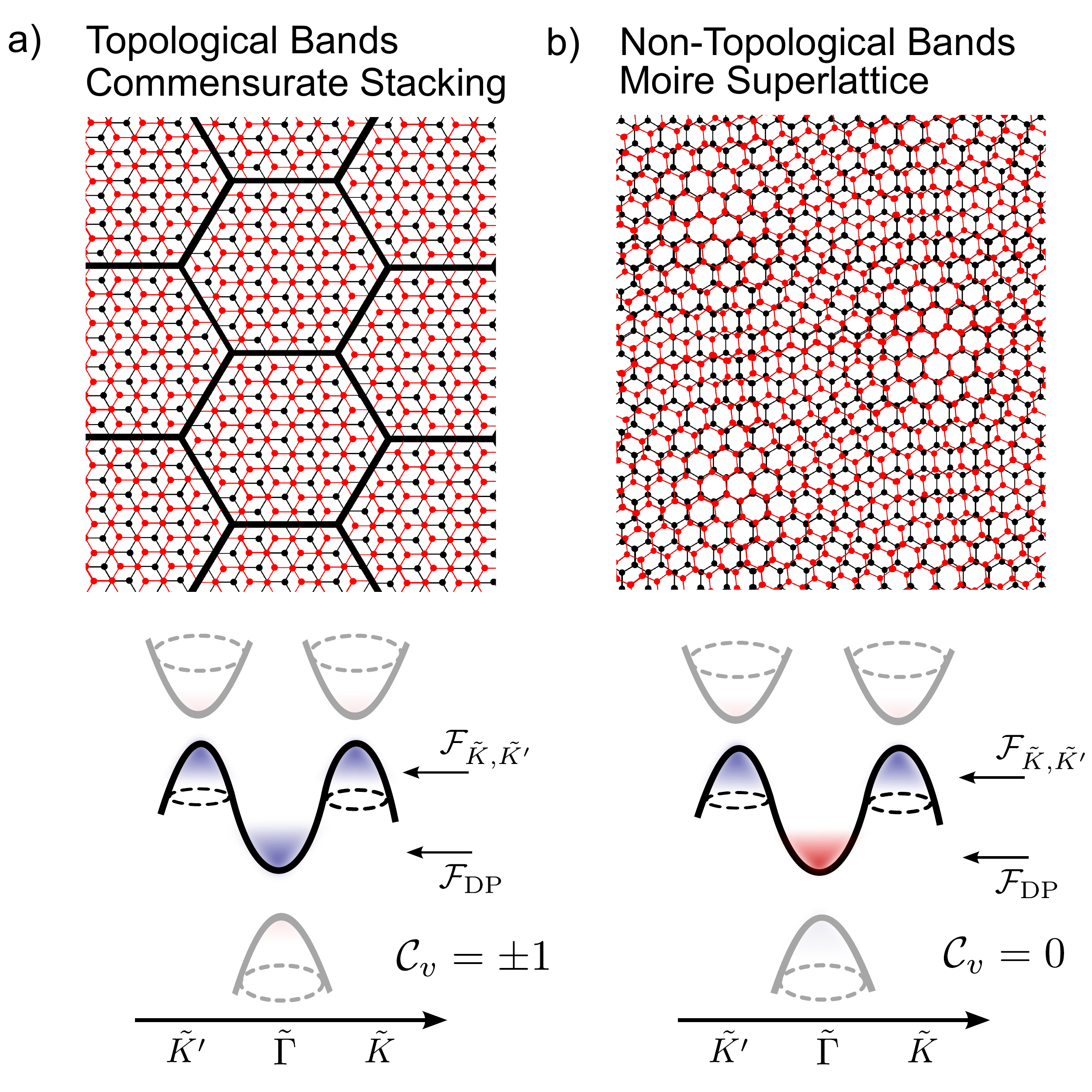}
\caption{ Topology of Bloch bands for different stacking types of G/$h$BN, commensurate (a) and incommensurate (b). {\it Top panels:} hexagonal commensurate domains (black lines mark domain walls) and  incommensurate Moir\'e superlattice structure. {\it Bottom panels:} valley Chern number $\mathcal{C}_v=\pm 1$
and  $\mathcal{C}_v=0$ for the lowest reconstructed minibands labeled ``1'' in Fig.\,\ref{fig2} (see text). This corresponds to the contributions to the net Berry flux, $\mathcal{F}=\mathcal{F}_{\rm DP}+\mathcal{F}_{\tilde K,\tilde K'}$, from the superlattice Brillouin zone center $\tilde{\Gamma}$ and corners $\tilde{K}$, $\tilde{K'}$
that have equal signs and opposite signs, respectively. Red and blue shaded regions indicate the $+$ and $-$ signs of Berry curvature. }
\label{fig1}
\vspace{-8mm}
\end{figure}

Here, we lay out an approach for engineering designer topological materials out of stacks of generic materials -- ``Chernburgers''. 
Our scheme naturally produces (i) topological bands with different Chern invariant values,  and (ii) tunable topological transitions.
As an illustration, we analyze Graphene on hexagonal Boron Nitride heterostructures (G/$h$BN), where broken inversion symmetry is expected to generate Berry curvature\cite{vhe,mak_2014} -- a key ingredient of topological materials. Indeed, recently valley currents have been demonstrated in G/$h$BN \cite{gorbachev14} signaling the presence of Berry curvature \cite{vhe}. As we will show, Berry curvature in G/$h$BN can be molded by stacking configuration, leading to a large variability in properties. Transitions between different topological states can be induced by a slight change in stacking angle.

\begin{figure}[t]
\includegraphics[scale=0.3]{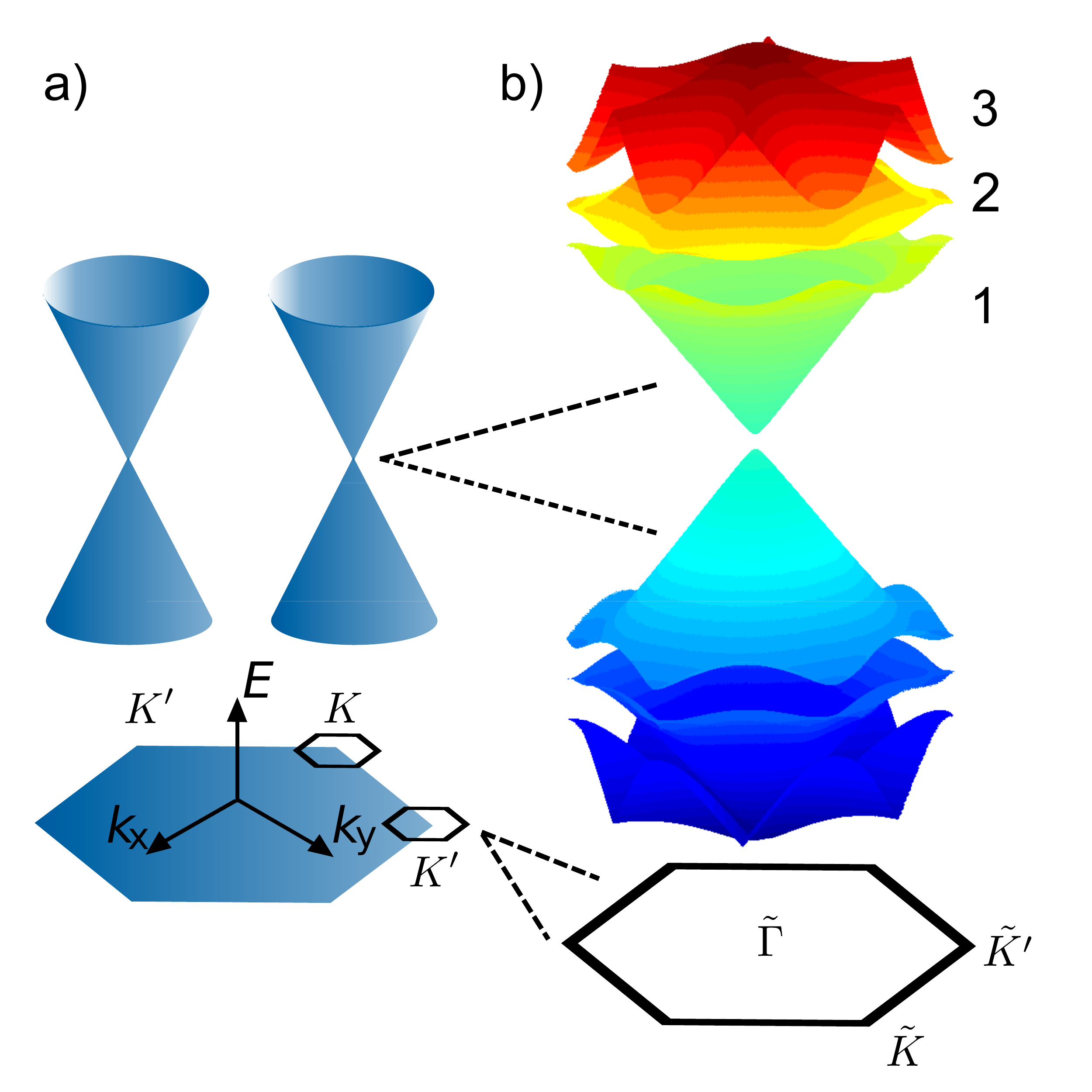}
\caption{Graphene superlattice potential transforms the massless Dirac bandstructure near points $K$, $K'$ of 
graphene Brillouin zone (a) into a family of minibands (b). Shown is the bandstructure near points $K$, $K'$ obtained from the Hamiltonian in Eq.\eqref{eq:hamiltonian} [parameters used: $m_3 =  60\,{\rm meV}$, $\epsilon_0= \hbar v|\vec b| = 1\,{\rm eV}$, $\Delta_{\rm g} = 20\,{\rm meV}$]. Large superlattice 
period 
translates into a small size of superlattice Brillouin zones (two hexagons positioned at points $K$, $K'$). 
Sublattice A/B dependent coupling [$\sigma_3$ term in Eq.\eqref{eq:hamiltonian}] generates a Dirac mass term and opens a gap between the conduction and valence bands; it also creates avoided band crossings 1-2-3 
above and below Dirac points.}
\label{fig2}
\vspace{-5mm}
\end{figure}

Topological bands in G/$h$BN arise separately for valley $K$ and valley $K'$.
Graphene bandstructure reconstruction due to the coupling to $h$BN produces superlattice minibands  \cite{Park2008_NatPhys,Park2008_PRL,guinea2010,kindermann2012,wallbank,song12}, with Berry curvature $\Omega(\vec k)$ developing near avoided crossings. The minibands for each valley possess a valley Chern number
\be
\mathcal{C}_v = \frac1{2\pi} \int_{k\in {\rm SBZ}} d^2k \Omega(\vec k)
,
\label{eq:valleychern}
\ee
where the integral is taken over the entire superlattice Brillouin zone (SBZ) in one valley ($K$ or $K'$). As discussed below, for commensurate stackings (Fig. 1a) 
$\mathcal{C}_v=\pm 1$ for the lowest minibands. 
In contrast, for incommensurate Moir\'e superlattice structures (Fig. 1b), the invariant (\ref{eq:valleychern}) vanishes in these minibands, $\mathcal{C}_v =0$. The difference in the behavior for these
configurations arises from the difference in sign of Berry flux contributions from regions near SBZ center  $\tilde\Gamma$ 
(the Dirac point, hereafter denoted DP)
and corners $\tilde{K}$, $ \tilde{K'}$ (see Fig.\,\ref{fig2}).  We will see that these contributions add in the commensurate case 
but subtract for the incommensurate case, yielding topological and non-topological bands, respectively (Fig. 1).

Interestingly, the conditions for both topological and non-topological bands are met by currently available systems. Indeed, both commensurate 
and incommensurate stackings have been recently identified in G/$h$BN by scanning probe microscopy\cite{leroy,woods14}. Further, the commensurate-incommensurate transition can be controlled by twist angle between G and $h$BN, providing a practical route in which to tailor electron topology via a tunable structural transition. 

We note that time reversal (TR) symmetry requires that $\Omega(\vec k)$ in $K$ and $K'$ valleys have opposite signs. As a result, the total Chern invariant always vanishes,
$\mathcal{C}_v(K)+ \mathcal{C}_v(K') = 0$. However, the weakness of inter-valley scattering \cite{gorbachev07,morpugolong} 
can enable long-range topological currents in individual valleys. As we will see, the non-local electrical signals mediated by such currents can provide diagnostics for valley band topology.

We also note that topological bands in graphene are sometimes presumed either impossible or impractical. Indeed, a connection between $K$ and $K'$ bands at high energies, whenever present, renders valley-specific topological invariants ill-defined\cite{ivar,jeiljung}. 
Proposals relying on large spin-orbit coupling  \cite{kanemele,weeks2011} are also sensitive to disorder;
proposals in other systems such as optical flux lattices \cite{cooper11} suffer from similar implementation pitfalls. Our scheme circumvents these difficulties by exploiting Bragg scattering in the G/$h$BN superlattice to create energy gaps above and below the $K$ and $K'$ Dirac points (see Fig.\,\ref{fig2}). The Dirac
points, sandwiched between these gaps, are no longer connected in a single band; the resulting minibands possess well-defined topological invariants.

\section{Minimal model for superlattice bands}

Modeling the superlattice bandstructure is greatly facilitated by several aspects of the G/$h$BN system. First is the long-wavelength superlattice periodicity, which results from nearly identical periods of graphene and $h$BN crystal structure.  For commensurate stackings, the superlattice structure is defined by a periodic array of hexagonal domains (Fig.\,\ref{fig1}a). Its periodicity, which is set by  the size of the domains, is on the order of  $\lambda\approx 100$ atomic distances. Likewise, in incommensurate stackings (Fig.\,\ref{fig1}b) the lattice mismatch and the twist angle between graphene and $h$BN produce long-period Moir\'e patterns with wavelength $\lambda\approx 10\,{\rm nm}$. Since the corresponding superlattice wavevector $b=2\pi/\lambda$ is too small to produce Bragg scattering between valleys $K$ and $K'$, the bandstructure reconstruction can be understood in terms of a Dirac model, giving a SBZ mini-bandstructure  separately for the $K$ and $K'$ valleys (see Fig.\,\ref{fig2})
\cite{Park2008_NatPhys,Park2008_PRL,guinea2010,kindermann2012,wallbank,song12,guineabandstructure,jeiljungbandstructure}.

Another property of the G/$h$BN system that simplifies modeling is a relatively weak coupling strength. Indeed, the reported values for the $h$BN-induced energy gap at the Dirac point are of order $500\,{\rm K}$\cite{hunt13,woods14,gorbachev14}, which is about 10 times smaller than the energy $\epsilon_0=\hbar v b$, where $b$ is the superlattice wavevector and $v=10^6\,{\rm m/s}$. This disparity  allows one to describe the superlattice bandstructure, for both commensurate and incommensurate stackings, with the effective Hamiltonian 
\be
H = v {\boldsymbol \sigma} \cdot \vec{p} + \Delta(\vec r) \sigma_3, \quad \Delta (\vec r)  = \Delta_{\rm g} + m_3 \sum_{j=1}^{3} {\rm cos}( \vec{b}_j\cdot \vec r) 
\label{eq:hamiltonian}
\ee
where $\Delta(\vec r)$ accounts for the coupling between graphene and $h$BN. Our minimal model for  $\Delta(\vec r)$, given in Eq.\eqref{eq:hamiltonian}, is sufficient to understand the key features of the bandstructure for both stacking types. In particular, $\Delta_{\rm g}$ describes the global gap at DP (point $\tilde{\Gamma}$), whereas $m_3$ describes Bragg scattering that creates avoided crossings at $\tilde{K}$ and $\tilde{K'}$ (see Fig.\,\ref{fig2}). As a result, the minibands are distinct and separated by energy gaps disconnecting the original $K$ and $K'$ points.
Our microscopic analysis, presented below, indicates that the terms $\Delta_{\rm g}$ and $m_3$ are present for both commensurate and incommensurate stackings. Crucially, the two cases are distinguished by opposite relative signs of $\Delta_{\rm g}$ and $m_3$.  This sign difference, as we will see, is key in producing different topological classes. 

We note parenthetically that a more general Hamiltonian can also include a scalar potential term modulated in the same way as the $\sigma_3$ term above \cite{Park2008_NatPhys,Park2008_PRL,wallbank}. However, as discussed elsewhere\cite{song12} the electron interaction effects strongly enhance the $\sigma_3$ coupling, but leave the scalar potential unrenormalized. Hence, we take $\sigma_3$ term as the dominant part of superlattice potential ignoring the scalar potential contribution. On similar grounds we disregard possible modulation of $\sigma_1$ and $\sigma_2$ types that may arise due to strain. 

\section{Global gap and the signs of $\Delta_{\rm g}$ and $m_3$}

Turning to the analysis of the coupling in Eq.\eqref{eq:hamiltonian}, 
we first consider the commensurate case, where all the hexagonal domains adopt the same lowest energy atomic configuration.
The simplest arrangement to produce such a stacking is perfect crystal axes alignment when $G$ and $h$BN lattices conform with each other as pictured in Fig.\,\ref{fig1}a.\footnote{While we have used AB stacking where G and $h$BN crystal axes are aligned in our illustration in Fig.\,\ref{fig1}a, other stackings can be also used and do not affect the conclusions in this work.}  Other commensurate stackings in the absence of perfect crystal axes alignment may also occur and do not affect our main conclusions. The registration within each hexagonal cell is locked producing an A/B sublattice asymmetry in graphene. Crucially, the sign of this asymmetry cannot change upon lateral sliding which is not accompanied by a rotation. Hence the asymmetry is of the same sign throughout the structure, leading to a global constant-sign gap.

To illustrate this important point, we present the argument in a form that does not depend on detailed knowledge of the registration within each of the domains. Of course, in practice the registration types (and hence the asymmetry signs) arise from general energetic and geometric constraints which can be easily accounted for\cite{guinea}. As an example, we consider three possible registrations:\\ (i) site A in $h$BN aligned with site A in graphene and site B in $h$BN with site B in graphene;\\ (ii) site A in $h$BN aligned with site B in graphene and site B in $h$BN with H (hollow) in graphene; \\ (iii) A in $h$BN aligned with site B in graphene whereas site B in $h$BN aligned with site A in graphene.
\\ 
Configurations (i) and (iii) cost the same energy, but have a different energy than (ii). Importantly, lateral sliding of a cell with configuration (i) cannot generate configuration (iii) since it would require a lattice rotation. Similarly, while lateral sliding of a cell with configuration 
(i) can generate configuration 
(ii), it costs a different energy. As a result, stacking frustration between neighboring cells cannot occur, locking the registration between all hexagonal cells to yield a constant global gap, $\Delta_{\rm g}$. 

Next, we note that imperfect registration around the domain boundaries
yields a weaker coupling between G and $h$BN (strained graphene sheet buckles \cite{woods14} increasing the G-to-$h$BN distance). 
Reduction in sublattice-asymmetric potential $\Delta_{\rm g,0}$ can be modeled as
\be
\Delta(\vec r) = \Delta_{\rm g,0} + \delta m [G(\vec r) \ast F(\vec r)],  \quad {\rm sgn}\,\delta m = - {\rm sgn}\, \Delta_{\rm g,0}
\ee
where $F(\vec r)$ describes the unit cell of the pattern of domain walls, $G(\vec r) = \sum_{n, l \in \mathfrak{Z}} \delta( \vec r - n\vec a_1 - l \vec a_2)$ is the superlattice form factor ($\vec a_{1,2}$ are superlattice basis vectors), and 
$\ast$ indicates convolution. The relative sign 
${\rm sgn}\,\delta m = - {\rm sgn} \,\Delta_{\rm g}$ accounts for the weaker coupling between G and $h$BN at the domain boundaries. 

Since we are interested in bandstructure reconstruction in the lowest minibands, we expand $\Delta(\vec r)$ into lowest harmonics yielding 
Eq.\eqref{eq:hamiltonian} with 
\be
\Delta_{\rm g} = \Delta_{\rm g,0} + \delta m \tilde{F}_{\vec q=0}, \quad m_3=  2 \delta m \tilde{F}_{\vec q = \vec b_j} 
\ee
where $\tilde{F}(\vec q) =  \frac{1}{\cal A}\int d^2 \vec r F(\vec r) e^{i \vec q \cdot \vec r}$ is the form factor, $\vec b_j$ are the reciprocal superlattice vectors, and ${\cal A}$ is the area of  
superlattice unit cell. Crucially the sign of the form factor $F$ determines the sign of $m_3$. Choosing a symmetric $F(\vec r)$, with origin at the centre of a hexagonal domain (pictured in Fig.\,1a, $\hat{\vec x}$ and $\hat{\vec y}$ are the horizontal and vertical directions) and $\delta$-functions along the hexagonal domain walls, we obtain the form factor
\be
\tilde{F}(\vec q) = \frac{2w}{\cal A}\sum_{j=1}^{3}  \frac{{\rm sin} (\frac{d}{2} \vec q_i \cdot \hat{\vec x})}{\vec q_i \cdot \hat{\vec x}} {\rm cos} \left(\frac{\sqrt{3}d}{2} \vec q_i \cdot \hat{\vec y}\right)
.
\label{eq:Fofq}
\ee
Here $d$ and $w$ are the domain wall length and width, and
$\vec q_i =  R(\theta_j) \vec q $, where $R(\theta_j)$ are the $2\times 2$ rotation matrices with $\theta_1 =0$, $\theta_2 = \pi/3$, and $\theta_3 = 2\pi/3$. 

\begin{figure}[t]
\includegraphics[width=0.5\textwidth]{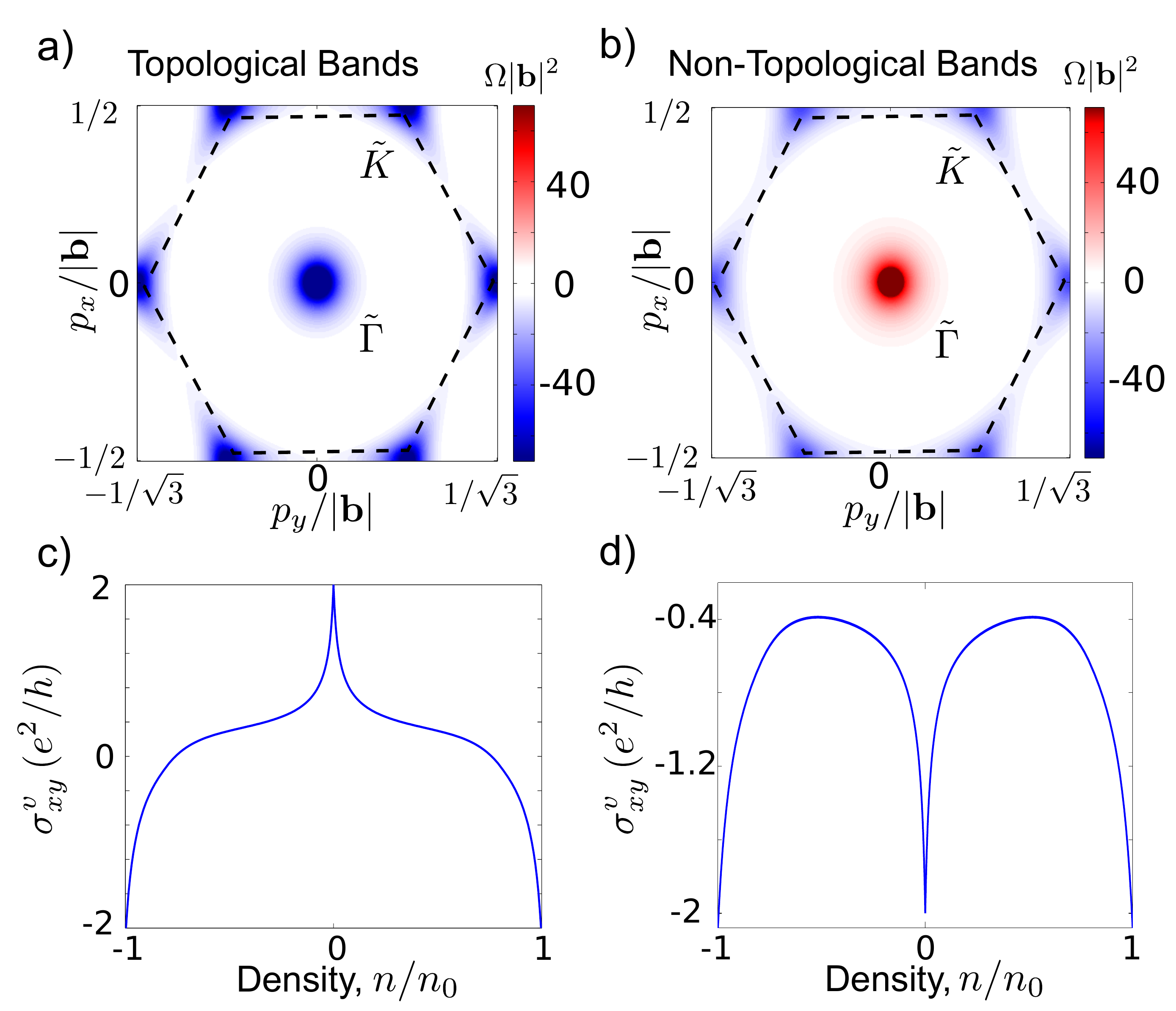}
\caption{ 
a,b) Berry curvature distribution, $\Omega(\vec k)$, in the lowest conduction band (labeled ``1'' in Fig.\,\ref{fig2}b) obtained from the Hamiltonian in Eq.\eqref{eq:hamiltonian}. Two choices of signs, 
(a) ${\rm sgn} (\Delta_{\rm g}) = {\rm sgn} (m_3)$ and (b) ${\rm sgn} (\Delta_{\rm g}) =  - {\rm sgn} (m_3)$, yield $\mathcal{C}_v = -1$ and $\mathcal{C}_v=0$ respectively. The hot spots of $\Omega(\vec k)$ at DP (point $\tilde\Gamma$) and SBZ corners $\tilde K$, $\tilde K'$ correspond to gap opening and avoided band crossing regions.
The central peak carries a net Berry flux $\pi$, whereas the corner peaks carry a net flux $\pm\pi/2$, see main text. 
Parameters used: $m_3 = 20 \, {\rm meV}$, $\epsilon_0=\hbar v |\vec b| = 300 \, {\rm meV}$, $\Delta_{\rm g} = m_3$ in (a), $\Delta_{\rm g} = -m_3$ in (b).
c, d) Valley Hall Conductivity, $\sigma_{xy}^v$ [Eq.\eqref{eq:valleyhall}], {\it vs.} carrier density for the two minibands above and below DP:
$\sigma_{xy}^v$ changes sign for topological bands (panel c) but keeps the same sign for non-topological bands (panel d)  [$n_0$ is the density needed to fill the first miniband, other parameter values same as in (a,b)]. 
}
\label{fig3}
\end{figure}

Evaluating Eq.\eqref{eq:Fofq} gives $\tilde{F}_{\vec q=0} = 3 \zeta>0$ and $\tilde{F}_{\vec q = \vec b_j}  = - 9\sqrt{3}\zeta /4\pi <0$, where $\zeta = wd/{\cal A} > 0$. Comparing with Eq.\eqref{eq:hamiltonian}, we find the relation between signs of $m_3$ and $\Delta_{\rm g}$:
\be\label{eq:signs_comm}
\sgn{m_3} = - {\rm sgn} (\delta m) = {\rm sgn} (\Delta_{\rm g})
.
\ee 
As we will see, this leads to a nontrivial topological class $\mathcal{C}_v = \pm 1$ in the lowest minibands (see Fig.\,\ref{fig3}a) .

The  incommensurate case (Moir\'e superlattice) differs from the commensurate case in two important ways. One is that the G-to-$h$BN coupling is dominated by the modulational part $\Delta(\vec r) = m_3 \sum_{j=1}^{3} {\rm cos}( \vec{b}_j\cdot \vec r) $ arising from the Moir\'e pattern. The other is that the global gap parameter  $\Delta_{\rm g}$ is zero in the bare Hamiltonian, however a nonzero $\Delta_{\rm g}$ value is generated perturbatively in $m_3$, with the $\Delta_{\rm g}$ sign the opposite of the $m_3$ sign. The analysis is particularly simple for the long-period Moir\'e patterns arising for rigid G and $h$BN stackings at small twist angles, as shown in Fig. 1b. 

Of course,  one $m_3$ harmonic cannot produce 
an average global gap at DP since it is sign-changing, $\la e^{i\vec b\vec x}\ra=0$. 
However, a combination of three different harmonics can open up a gap \cite{song12}. This can be seen from a perturbation analysis of the Hamiltonian \eqref{eq:hamiltonian} which we write as $H=H_0+V$, where $H_0=v{\boldsymbol\sigma}\cdot\vec p$, $V=\sigma_3 m_3\sum_{j=1}^3\cos(\vec b_j\cdot\vec r)$. Perturbation theory in $V$ yields a term describing a global gap at a third order in $V$ via 
\be
\delta H=V\frac1{\epsilon-H_0}V\frac1{\epsilon-H_0}V
.
\ee 
Choosing triplets of harmonics with $\vec b_i +\vec b_j +\vec b_k = 0$, third-order perturbation theory in $m_3$ yields a gap 
\be\label{eq:Delta_0^0}
\Delta_{\rm g}=\!\!\!\!\sum_{\pm\vec b_i,\pm\vec b_k}\!\!\! \frac{m_3\sigma_3}{2}\frac1{v\boldsymbol{\sigma}\cdot\vec b_i}\frac{m_3\sigma_3}{2}\frac1{v\boldsymbol{\sigma}\cdot\vec b_k}\frac{m_3}{2}=-\frac{3m_3^3}{4(v|\vec b|)^2}
,
\ee
where the minus sign results from the anticommutation relations $[ \sigma_{1},\sigma_3]_+ = 0$, $[ \sigma_{2},\sigma_3]_+ = 0$. Importantly, this analysis predicts a relation between signs
\be\label{eq:signs_incomm}
{\rm sgn} (\Delta_{\rm g}) = - {\rm sgn} (m_3)
\ee
which is opposite to the relation found for the commensurate case, Eq.\eqref{eq:signs_comm}. While the gap size obtained at a third order of perturbation theory in a non-interacting system is small, electron interaction effects are expected to produce an enhancement and generate a large $\Delta_{\rm g}$ \cite{song12}. As we will see, the signs in Eq.\eqref{eq:signs_incomm} lead to trivial topological classes for superlattice bands, $\mathcal{C}_v=0$ (see Fig.\,\ref{fig3}b). 

In addition to the difference in signs, the commensurate and incommensurate stackings differ in the relative magnitude of the $\Delta_{\rm g}$ and $m_3$ couplings. As we argued above, the global gap coupling $\Delta_{\rm g}$ dominates in the commensurate case, with a relatively weaker modulational part $m_3$ arising due to registration unzipping along domain boundaries. In contrast, the modulational coupling $m_3$ is dominant in the incommensurate case, with the global gap $\Delta_{\rm g}$ arising at third-order perturbation in $m_3$. The two distinct microscopic pictures result in a disparity between the $\Delta_{\rm g}$ and $m_3$ scales and a sign difference,  ultimately leading to different topological classes.

\section{Topological classes}

We proceed to explore how stacking types impact the band topology. 
The topological properties of G/$h$BN can be analyzed through the Berry curvature in the minibands. Even though the G/h-BN Hamiltonian, Eq.\eqref{eq:hamiltonian}, possesses TR symmetry, its broken inversion symmetry allows for a finite Berry curvature to develop in the SBZ 
\be
\Omega_n(\vec k) = \nabla_\vec{k} \times \vec{A}_n(\vec k), \quad \vec A_n(\vec k) = i \bra{u_n(\vec{k})} \nabla_\vec{k} \ket{u_n(\vec k)}
\label{eq:berryconnection}
.
\ee
Here $n$ is the band index, $\vec{A}$ is the Berry's connection, and $\ket{u_n(\vec k)}$ are the eigenvectors of Eq.\eqref{eq:hamiltonian}. 
In what follows, we concentrate on a single valley and the lowest conduction mini-band (labeled ``1'' in Fig.\,\ref{fig2}b).

Using Eq.\eqref{eq:berryconnection}, we evaluate $\Omega_n(\vec k) $ for the bandstructure generated by Eq.\eqref{eq:hamiltonian}, and obtain Berry curvature maps in SBZ reciprocal space which are shown in Fig.\,\ref{fig3}a,b. We adopted a numerical method similar to that outlined in Ref. \cite{japphys}, see supplement for a full description \cite{supplement}. In Fig.\,\ref{fig3}a,b we plot $\Omega(\vec k) $ corresponding to the lowest conduction band (labeled ``1'' in Fig.\,\ref{fig2}b); the lowest valence band exhibits the same behavior but with opposite sign.
We find that $\Omega(\vec k)$ is concentrated in the reciprocal space regions where the bandstructure exhibits gaps and avoided crossings, namely at the Dirac points and SBZ corners [$\tilde\Gamma$, and $\tilde K$, $\tilde K'$ respectively]. 

Integrating $\Omega(\vec k) $ over the superlattice Brillouin zone to obtain the valley Chern number, Eq.\eqref{eq:valleychern}, we identify two distinct cases. For the equal-sign case, Eq.\eqref{eq:signs_comm}, which corresponds to commensurate stackings, we obtain $\mathcal{C}_v = -1$ (see Fig.\,\ref{fig3}a). For the opposite-sign case, Eq.\eqref{eq:signs_incomm}, which corresponds to incommensurate stackings, 
we obtain $\mathcal{C}_v = 0$ (see Fig.\,\ref{fig3}b). This behavior can be understood in terms of the Berry curvature hot spots in SBZ (see Fig.\,\ref{fig3}). In particular, 
when $\Omega(\vec k) $ at DP ($\tilde\Gamma$) as well as $\tilde K$ and $\tilde K'$ are of the same sign and add to give $\mathcal{C}_v = -1$, a topological band is obtained (Fig.\,\ref{fig3}a). On the other hand, if $\Omega(\vec k) $ sign at DP is opposite to that at $\tilde K$ and $\tilde K'$, Berry fluxes subtract giving $\mathcal{C}_v = 0$. In this case, a non-topological band is obtained (Fig.\,\ref{fig3}b).  

To gain more insight into band topology in Fig.\,\ref{fig3}, it is instructive to analyze the hot spots of $\Omega(\vec k)$. 
Near SBZ center $\tilde\Gamma$, the bandstructure is approximated by a constant-mass Dirac Hamiltonian
$ 
H= v \sigma \cdot \vec{p} + \Delta_{\rm g} \sigma_3
$,
where $\Delta _0 \equiv \Delta_K=\Delta_{K'}$ (due to TR Symmetry). Berry curvature is then given by the well-known expression
\be
\Omega_{\pm,K(K')} (\vec k) = \mp  \frac{\Delta_{\rm g} v^2 \eta_z}{2(v^2\vec p^2+\Delta_{\rm g}^2)^{3/2}},
\ee
where $\pm$ refer to the conduction and valence bands and $\eta_{z} = +$ and $\eta_{z} = -$ for valley $K$ and $K'$ 
respectively. This translates into the net Berry curvature flux which is controlled by the sign of $\Delta_{\rm g}$:
\be
\mathcal{F}_{\rm DP} = \int d^2k \Omega_{\pm ,K(K')} (\vec k) =  \mp \pi \eta_z {\rm sgn}(\Delta_{\rm g})
\label{eq:dpflux}
\ee
giving $\mp\pi$ for $K$, $K'$ valleys, as expected for a Dirac point. 

As illustrated in Fig.\,\ref{fig3}, Berry curvature also features hot spots at SBZ corners $\tilde K$ and $\tilde K'$. These arise from Bragg scattering by the superlattice harmonics in Eq.\eqref{eq:hamiltonian} which mix the pseudospin textures; the energy spectrum and $\Omega(\vec k)$ close to $\tilde K, \tilde K'$ can be modeled using the $\vec k \cdot \vec p$ method, see supplement \cite{supplement}. We find that the net Berry flux in the conduction band, $\mathcal{F}_{\tilde K, \tilde K'} = \int d^2k \Omega(\vec k)$ about the corners of the SBZ are controlled by $m_3$,
\be
{\cal F}_{\tilde K, \tilde K'} =  - \frac{\pi}{2} \eta_z {\rm sgn}(m_3),
\label{eq:ktildeflux}
\ee
and are equal for both $\tilde K$ and $\tilde K'$. When $m_3$ 
becomes small the hot spots
around $\tilde K$, $\tilde K'$ contract, however the net flux $\pm\pi/2$ for each hot spot remains unchanged. 

We note that the ``half-Dirac" flux $ \pm \pi/2$ follows from Chern number quantization.
Integer ${\cal C}_v = \frac1{2\pi} \int_{k\in SBZ} d^2k \Omega(\vec k)$ 
arises from summing the Berry curvature concentrated about DP and $\tilde{K}, \tilde{K'}$ points in the SBZ (as shown in Fig.\,\ref{fig3}a,b). Since there are two inequivalent $\tilde{K}$ points in the SBZ, $\mathcal{C}_{v} = \frac{1}{2\pi} \Big(\mathcal{F}_{\rm DP} + 2 \mathcal{F}_{\tilde{K},\tilde{K'}}\Big)$. 
Integer $\mathcal{C}_v$ and $\mathcal{F}_{\rm DP}=\pm\pi$ yield $\pm \pi/2$ values for $\mathcal{F}_{\tilde{K},\tilde{K'}}$.

\begin{figure}[t]
\includegraphics[width=0.5\textwidth]{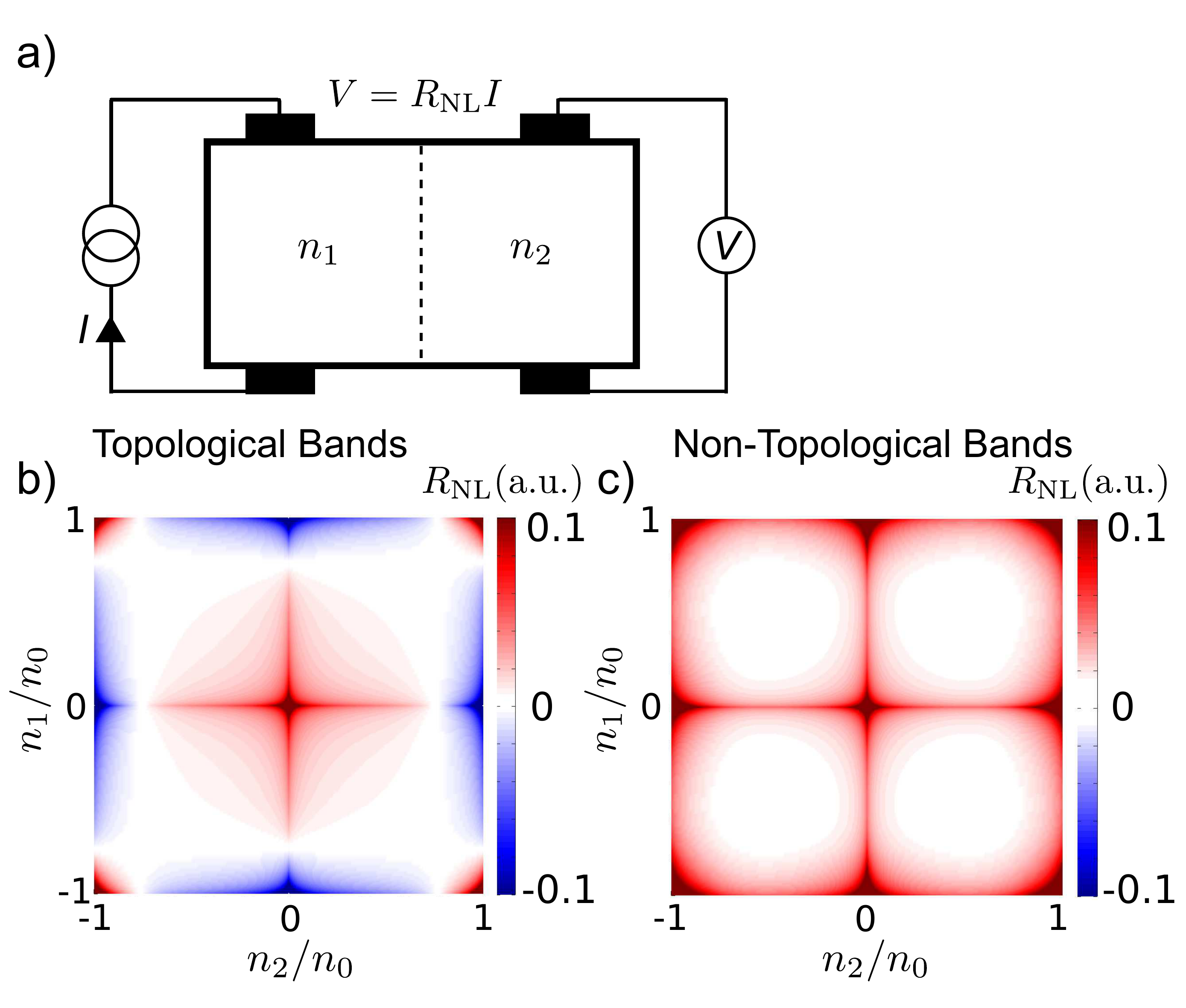}
\caption{ a) Nonlocal response as an all-electrical diagnostic of the Berry curvature energy dependence and of valley Chern numbers $\mathcal{C}_v$. Shown schematically is an H-geometry 
with separately gated injection and detection regions allowing carrier density $n_1$, $n_2$ in these regions to be  tuned independently (see text). b,c) The nonlocal resitance $R_{\rm NL} = A \sigma_{xy}^v (n_1) \times  \sigma_{xy}^v(n_2)$ features multiple sign changes as a function of $n_1$, $n_2$ for topological bands and no sign changes for non-topological bands.  Parameters used are the same as in Fig.\,\ref{fig3}a,b; the corresponding dependence $\sigma_{xy}^v$ vs. $n$ is shown in Fig.\,\ref{fig3}c,d.
}
\label{fig4}
\end{figure}

\section{Valley currents and Berry curvature spectroscopy}

Topological currents associated with each of the valleys can propagate over extended distances so long as the inter-valley scattering is weak\cite{gorbachev07}. While TR symmetry requires no net charge Hall currents, the opposite signs of $\Omega(\vec k)$ in $K$ and $K'$ allow transverse valley currents, $\vec J_v = \vec J_{K} - \vec J_{K'}$, to be induced by a longitudinal Electric field, $\vec E$. This valley Hall effect (VHE) is described by \cite{vhe}
\be
\vec J_v = \sigma_{xy}^v \vec E \times \hat{\vec n}, \quad \sigma_{xy}^v  =  \frac{Ne^2}{h} \int \frac{d^2 k}{2\pi} \Omega(\vec k) f(\vec k)
\label{eq:valleyhall}
\ee 
where $\hat{\vec n}$ points perpendicular to G/$h$BN, $N=4$ is valley/spin degeneracy, and $f(\vec k) = (e^{\beta (\epsilon_k - \mu)} +1)^{-1}$. 

The difference between topological bands and non-topological bands is 
reflected in the behavior of $\sigma_{xy}^v$ which changes signs as a function of density 
varying  in a single band, or maintains a constant sign, as illustrated in Fig.\,\ref{fig3}c,d.
We note that sign-changing $\sigma_{xy}^v$ does not contradict $\mathcal{C}_v =\pm 1$ for topological bands, since $\mathcal{C}_v$ tracks the total change in $\sigma_{xy}^v$ as density is swept through the band. Accounting for a total change of $\sigma_{xy}^v$ (quantized for topological bands, vanishing for non-topological bands), and using $\sigma_{xy}^v$ exactly at $\tilde{\Gamma}$ with magnitude $N e^2/2h$ obtained from counting net Berry flux\cite{ivar,jeiljung,lensky}, we obtain the contrasting $\sigma_{xy}^v$ shown in Fig.\,\ref{fig3}c,d.

Even though the currents $\vec J_v$ are chargeless, they can be detected by electrical means\cite{gorbachev14}. Indeed, the long propagation lengths enabled by weak inter-valley scattering allow valley currents flowing in system {\it  bulk} to mediate non-local electrical response. This is distinct from graphene edge modes that are highly susceptible to localization and gapping out on rough or imperfect edges. In contrast, recent measurements of inter-valley scattering in G/$h$BN yield mean free paths as large as several microns \cite{gorbachev07,morpugolong}. Non-local resistance measurements (Fig.\,\ref{fig4}) can therefore provide an all-electrical and robust way to probe the bulk valley-Hall conductivity.  

Non-local resistance, $R_{\rm NL}$ arises in a way illustrated in Fig.\,\ref{fig4}a. Transverse valley currents, $\vec J_v$, induced by an electrical current $I$ can propagate over extended distances to induce a valley imbalance profile across the device, $\delta \mu = \delta \mu_K - \delta \mu_{K'}$ . Even far away from the current source, valley imbalance $\delta \mu$ can set up an appreciable transverse electric field via the reverse valley Hall effect, 
\be
\vec E =  \frac{\sigma_{xy}^v(n_2)}{\sigma_{xx} (n_2)} (\nabla \delta \mu) \times \hat{\vec n}.
\ee
This provides the key mechanism through which the chargeless long-range valley currents are converted to an electric signal at the readout contacts, producing a nonlocal trans-resistance $R_{\rm NL}$. 
Such $R_{\rm NL}$ was recently observed in Ref. \cite{gorbachev14} for a uniform density device $n_1 = n_2$. Importantly, control over local density in the geometry of Fig.\,\ref{fig4}a, $R_{\rm NL} = V/I =  A \sigma_{xy}^v (n_1) \times  \sigma_{xy}^v(n_2)$ is sensitive to the density and signs of $\sigma_{xy}^v$ in the two regions. Here the prefactor $A$ is positive and depends on the longitudinal conductivity $\sigma_{xx}$ of both regions, device dimensions, and intervalley scattering length, similar to that analyzed for the spin-Hall effect \cite{dima}. For illustration, we set $A=(h/e^2)^3$.

Since $\sigma_{xy}^v$ for topological bands (Fig.\,\ref{fig3}c) changes sign as density is swept in a single band, we find that $R_{\rm NL}$ displays multiple sign changes as a function of density in $n_1,n_2$ as shown in Fig.\,\ref{fig4}b. The sign-changing behavior of $R_{\rm NL}$ can be traced back to the finite value of $\mathcal{C}_v = \pm1$ for topological bands and a $\sigma_{xy}^v$ of $N e^2/2h$ at neutrality  $n_{1,2}=0$ (i.e. at DP) \cite{ivar,jeiljung,lensky}. In contrast, $R_{\rm NL}$ maintains a constant sign for non-topological bands, $\mathcal{C}_v = 0$, as shown in Fig.\,\ref{fig4}c. As a result, sign changes in $R_{\rm NL}$ provide a clear diagnostic for topological bands.

In summary, graphene superlattices provide a practical route to constructing topological bands out of generic materials, as illustrated via tunable electron band topology in commensurate/incommensurate stackings. While we focused on Berry curvature diagnostic, graphene superlattices afford a new and widely accessible setting in which to achieve a wide variety of topological behavior. For example, chiral edge states associated with the boundary between topological and non-topological states may be found amongst adjacent domains in single samples with spatially varying twist angles. A number of different systems can be used, including SiC where superlattice stackings have been observed,\cite{berger06,lanzara07} G/$h$BN \cite{gorbachev14,hunt13,woods14} and twisted bilayer graphene.\cite{evaandrei10, evaandrei11,javier12} The ease with which stacked G/$h$BN structures can be made,\cite{dean} and the robust bulk transport signatures of their topological character open the door to access and probe electronic band topology in designer topological materials.  

This work was supported by STC Center for Integrated Quantum Materials, NSF Grant No. DMR-1231319 and in part by the U. S. Army Research Laboratory and the U. S. Army Research Office through the Institute for Soldier Nanotechnologies, under contract number W911NF-13-D-0001. J.C.W.S acknowledges support from a Burke Fellowship at Caltech and the NSS program (Singapore).

\begin{appendix}

\section{SUPPLEMENTARY INFORMATION}

Herein we briefly describe the numerical method used to compute Berry curvature, and discuss a $\vec k \cdot p$ theory for SBZ $\tilde{K},\tilde{K'}$ points. 

\section{Computing Berry Curvature}
The topological properties of G/$h$-BN can be analyzed through the Berry curvature in the minibands. Even though the G/h-BN Hamiltonian, Eq.(2) of the main text, possesses TR symmetry, its broken inversion symmetry allows for a finite Berry curvature to develop in the SBZ 
\be
\Omega_n(\vec k) = \nabla_\vec{k} \times \vec{A}_n(\vec k), \quad \vec A_n(\vec k) = i \bra{u_n(\vec{k})} \nabla_\vec{k} \ket{u_n(\vec k)}
\label{eq:berryconnection}
.
\ee
Here $n$ is the band index, $\vec{A}$ is the Berry's connection, and $\ket{u_n(\vec k)}$ are the eigenvectors of Eq. (2) of the main text. 
In the main text, we concentrate on a single valley and the lowest conduction mini-band (labeled ``1'' in Fig. 2b of the main text). % in the SBZ). 

Here we comment on the procedure used to evaluate Berry curvature. We use the eigenvectors, $\ket{u_n(\vec k)}$,  obtained at each $\vec k$ from numerical diagonalization of the Hamiltonian, Eq.(2) of the main text [in doing so we use $(2m +1)^2 $ points in the extended superlattice Brillouin zone scheme, typically with $m=3$]. Next, we calculate $\Omega_n(\vec k)$ using a gauge invariant method similar to that used in Ref. 28 of the main text.
We summarize this method briefly: adopting a fine mesh of the Brillouin zone (e.g. in Fig.3 of the main text the mesh for the grid shown was $400 \times 460$), the Berry curvature  at each $\vec k$ can be obtained by numerically integrating the Berry's connection $\vec{A}(\vec k)$ in small loops around $\vec k$ (viz. Stokes' theorem).  For each loop, we first choose a set of eigenvectors around the loop and then calculate Berry's connection between points on the loop. This ensures that the arbitrary phases appear twice in the loop integral but with opposite sign. In this way, (numerical) problems with gauge choice are eliminated as arbitrary phases introduced through the numerical diagonalization are cancelled in the loop.

\section{$\vec k \cdot \vec p$ theory for SBZ $\tilde{K},\tilde{K'}$ points}

\begin{figure}
\includegraphics[width=0.5\textwidth]{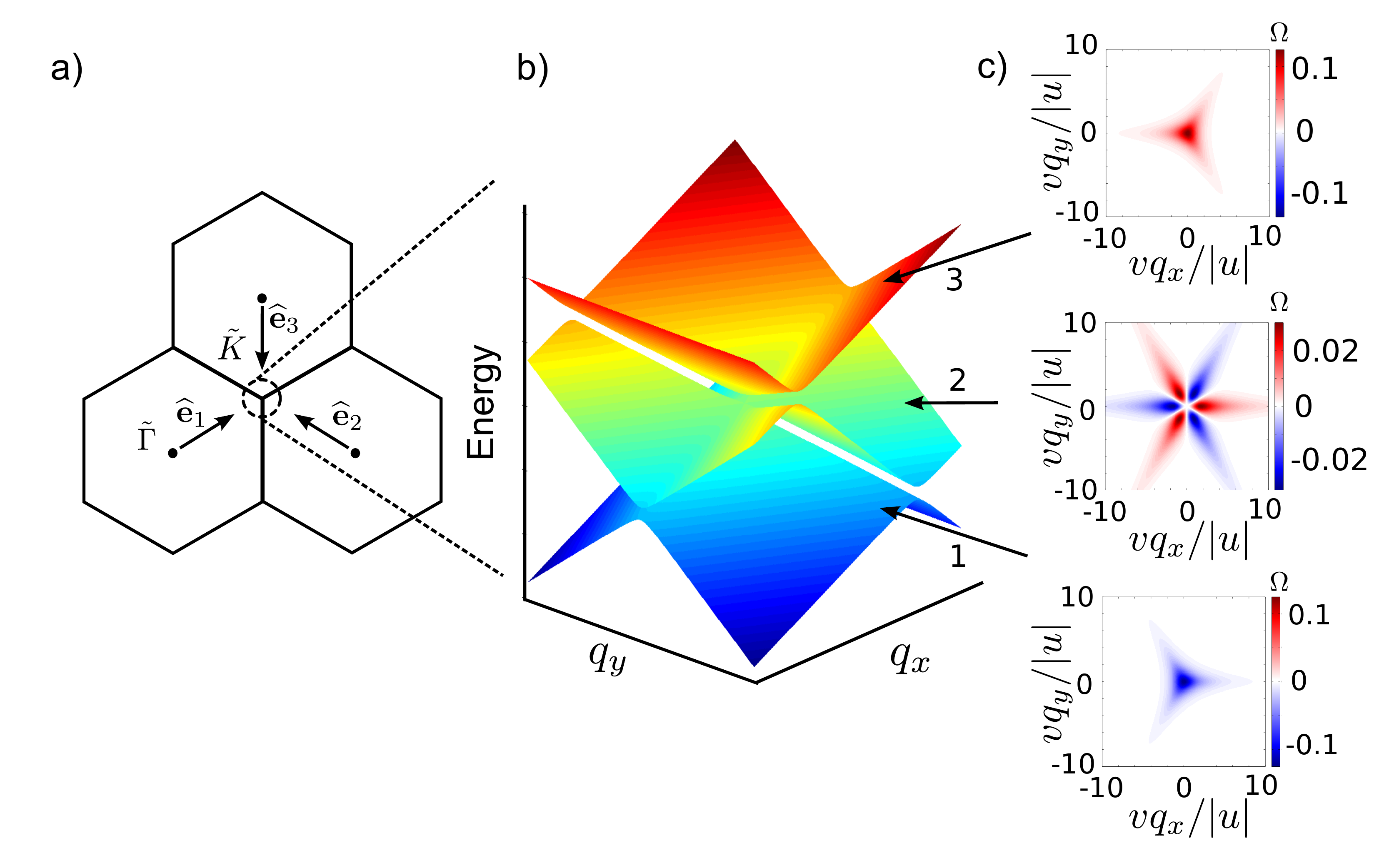}
\caption{Bandstructure and Berry curvature around $\tilde{K},\tilde{K'}$ points. a) Bragg scattering at SBZ corner points $\tilde K$, $\tilde K'$ mixes the three neighboring SBZs in the extended zone scheme. The avoided crossing of three bands at $\tilde{K}$, modeled by the $\vec k\cdot \vec p$ Hamiltonian, Eq.(\ref{eq:kp}), yields a bandstructure shown in (b) [energy and wavenumber units are $|u|$ and $|u|/v$]. c) Berry curvature distribution, $\Omega(\vec q)$, in reciprocal space features three-fold symmetry for bands 1 and 3 and six-fold symmetry for band 2. The net Berry curvature (Berry flux) is $-\pi/2$ and $\pi/2$ for bands 1 and 3, and zero for band 2. }
\label{fig2supp}
\end{figure}

The finite Berry curvature at $\tilde K$, $\tilde K'$ (in the SBZ) can be traced to mixing of the pseudospin textures by Bragg scattering shown in Fig. \ref{fig2supp}a. The bandstructure near $\tilde K$, $\tilde K'$ can be modeled using the $\vec k \cdot \vec p$ method. Both $\tilde K$ and $\tilde K'$ are points of triple band crossing in the limit of vanishing superlattice potential. 
We therefore introduce three states $|j\ra$, $j=1,2,3$, one for each crossing band. Using band dispersion linearized near $\tilde K$ (or, $\tilde K'$) and introducing a matrix element describing the coupling of the three states by Bragg scattering, we can write the $\vec k \cdot \vec p$ Hamiltonian in the conduction band as
\be \label{eq:kp}
H=\sum_{j=1,2,3} E_j(\vec p) |j\ra\la j|+\sum_{j'- j=1\,({\rm mod}\,3)} u|j'\ra\la j|
+{\rm h.c.}
\ee
where $E_j(\vec p)=\bra{\psi(\vec e_j)} v\sigma \cdot \vec p \ket{\psi(\vec e_j)}   = v\vec p\cdot \vec e_j$, and $u = u_{j,j'} = \bra{\psi(\vec e_j)} m \sigma_3 \ket{\psi(\vec e_{j'})} = i\frac{\sqrt{3}}{4} m_3 $. In the latter, $j\neq j'$ and are taken cyclically.
Here $\ket{\psi(\vec e_j)}$ is a spinor pointing along $\vec e_j$ and $\vec e_1$, $\vec e_2$, $\vec e_3$ are three unit vectors 
pictured in Fig.\ref{fig2supp}a. 
Evaluating the determinant in
\be
{\rm det}\,\lp\begin{array}{ccc} E_1(\vec p)-\epsilon & u & u^* \\ u^* & E_2(\vec p)-\epsilon & u \\ u & u^* & E_3(\vec p)-\epsilon \end{array}\rp =0
,
\ee
gives
\be
(E_1(\vec p)-\epsilon)(E_2(\vec p)-\epsilon)(E_3(\vec p)-\epsilon)+u^3+{u^*}^3+3|u|^2\epsilon=0
.
\ee
The resulting bandstructure is shown in Fig.\ref{fig2supp}b which mimics the behavior close to the $\tilde K, \tilde{K'}$ seen in Fig. 2b of the main text as expected. 

We use eigenstates obtained from the above Hamiltonian, Eq.(\ref{eq:kp}), to compute Berry curvature  as shown in Fig. \ref{fig2supp}c. Here we adopted the same numerical method as outlined in the above section. The Berry curvature distribution shown in Fig. \ref{fig2supp}c is concentrated close to the avoided crossings as expected from Fig. 3 of the main text. 

Summing up the Berry curvature in each of the bands we directly verify that bands 1 and 3 contribute $-\pi/2$ and $\pi/2$ to the net flux $\int d^2q \Omega(\vec q) $ respectively, whereas the net flux in band 2 is zero. In Fig. \ref{fig2supp} we used $m_3 >0$.  Flipping the sign of $m_3$ results in an opposite sign of flux in band 1 and 2 to that shown in Fig. \ref{fig2supp}c; band 2's net flux remains zero. As a result, we find that the net Berry flux in band 1 (the lowest conduction band) is described by Eq. (1) of the main text. 

Importantly, it is $m_3$ which controls the Bragg scattering at $\tilde K, \tilde{K'}$ that determines the Berry curvature at the edges of the SBZ. As discussed in the main text, the separate origins of Berry curvature at $\tilde \Gamma$ and $\tilde K, \tilde{K'}$ points allows for control over $\mathcal{C}_v$ of the reconstructed minibands of G/$h$-BN. 

\end{appendix}

\end{document}